\documentclass[preprint,a4paper,showpacs,preprintnumbers,superscriptaddress,amsmath,amssymb]{revtex4}
\pdfoutput=1
\usepackage{graphicx} 
\usepackage{dcolumn} 
\usepackage{bm} 

\begin{document}

\title{Non-invasive probing of persistent conductivity in high quality ZnCdSe / ZnSe quantum wells using surface acoustic waves}

\author{D. A. Fuhrmann}
\affiliation{Institut f{\"u}r Physik der Universit{\"a}t Augsburg, Experimentalphysik I,  Universit{\"a}tsstr. 1, 86159 Augsburg, Germany}
\author{H. J. Krenner}\email{hubert.krenner@physik.uni-augsburg.de}
\affiliation{Institut f{\"u}r Physik der Universit{\"a}t Augsburg, Experimentalphysik I,  Universit{\"a}tsstr. 1, 86159 Augsburg, Germany}
\author{A. Wixforth}
\affiliation{Institut f{\"u}r Physik der Universit{\"a}t Augsburg, Experimentalphysik I,  Universit{\"a}tsstr. 1, 86159 Augsburg, Germany}
\affiliation{Center for NanoScience (CeNS), Geschwister-Scholl-Platz 1, 80539 Munich, Germany}
\author{A. Curran}
\affiliation{School of Engineering and Physical Sciences, Heriot-Watt University, Edinburgh, EH14 4AS, United Kingdom}
\author{K. A. Prior}
\affiliation{School of Engineering and Physical Sciences, Heriot-Watt University, Edinburgh, EH14 4AS, United Kingdom}
\author{R. J. Warburton}
\affiliation{School of Engineering and Physical Sciences, Heriot-Watt University, Edinburgh, EH14 4AS, United Kingdom}
\affiliation{Department of Physics, University of Basel, Klingelbergstr. 82, 4056 Basel, Switzerland}
\author{J. Ebbecke}
\affiliation{Center for NanoScience (CeNS), Geschwister-Scholl-Platz 1, 80539 Munich, Germany}
\affiliation{Mads Clausen Institute, University of Southern Denmark, Sonderborg, Denmark}

\date{\today}

\begin{abstract}
Attenuation of a surface acoustic wave is used as a highly sensitive and non-invasive probe of persistent photoconductivity effects in ZnCdSe/ZnSe quantum wells. These effects are observed over long time-scales exceeding several minutes at low temperatures. By varying the optical excitation energy and power and temperature we show that these effects arise from carriers photogenerated by interband excitation which are trapped in random potential fluctuations in the quantum wells related to compositional fluctuations. Effects related to defect levels in the bandgap can be excluded and a transition of the conduction mechanism with temperature from a hopping to a percolation regime is observed. The transition temperature observed for our quantum well material is strongly reduced compared to bulk crystals. This indicates a superior structural quality giving rise to only weak potential fluctuation of $\lesssim 3 $ meV. 
\end{abstract}


\maketitle
\subsection{Introduction}
Surface acoustic waves (SAW) are a highly versatile tool to probe and manipulate optical and electronic properties of materials. Examples in the field of low-dimensional semiconductor heterostructures include charge conveyance and spectroscopy of high-mobility electron systems \cite{PhysRevB.40.7874,Rotter_PRL} and optically active systems as quantum wells (QWs) \cite{PhysRevB.57.R6850,couto_PRL}and quantum dots (QDs) \cite{boedefeld_PRB,gell_APL,couto_NatPhot}. However, most of these investigations focused on the most advanced III-V semiconductor compounds due to their piezoelectricity which allows for the use of interdigital transducers (IDTs) to excite and detect SAW. Despite the fact that most II-VI semiconductors are also weakly piezoelectric still SAW-based studies are very limited \cite{fuhrmann}.\\
One important mechanism which has large impact on the electrical transport properties in semiconductors is the persistent photoconductivity effect (PPC). This persistence of free charge carriers in semiconductors long after the removal of the exciting light source can by far exceed typical carrier lifetimes and leads to a pronounced change of conductivity in the material. PPC has been studied in the AlGaAs and other III-V material systems in great detail for many years \cite{Lang}. In the case of AlGaAs a thermal barrier prevents the recombination of the DX-centers at low temperatures which gives rise to the PPC effect. Similar PPC effects have also been observed in II-VI semiconductors such as ZnCdSe, ZnMgSSe, ZnMgSe and CdTe \cite{Firszt,Scholl1994491}. In inhomogeneous semiconductors \cite{Sheinkman76} and heterostructures \cite{Queisser} its origin can be of a different nature. In particular for bulk ZnCdSe spatial separation of charge carriers in random local potential fluctuations (RLPF) can occur. These RLPF result from compositional fluctuations during material growth and have been found to be the dominant mechanism giving rise to PPC \cite{PhysRevB.44.13343,Dissanayake,Jiang2,PhysRevLett.64.2547,Jiang1,PhysRevB.45.4520,PhysRevB.41.5178}. One striking consequence is the transition of the conduction mechanism from a localized to a percolation regime at a critical temperature. Surprisingly, a similar transition temperatures were found for both bulk and QWs grown by molecular beam epitaxy \cite{Chang} indicating that in these materials RLPFs are introduced inherently and cannot be reduced by introducing heterointerfaces.\\
In this paper we present a study of PPC in a ZnCdSe/ZnSe multiple quantum well (MQW) structure without directly measuring the conductivity by electrical means. Instead we apply a method based on surface acoustic SAWs, which for low SAW power represents a non-invasive and highly sensitive probe of the conductivity. The underlying mechanism is the attenuation of the SAW by mobile carriers which can be measured with high precision. Moreover, we find that for our QW material the transition temperature between the localized and percolation transport regime is significantly reduced to $T_C\lesssim 40$ K compared to values of $T_C^{bulk} \sim 120$ K reported for bulk material \cite{Jiang2}. This finding clearly shows the superior quality of our material compared to previous reports and indicates that inhomogeneities resulting in RLPFs are substantially reduced.\\ 

\subsection{Experimental setup}
\begin{figure}
	\includegraphics[width=0.9\columnwidth]{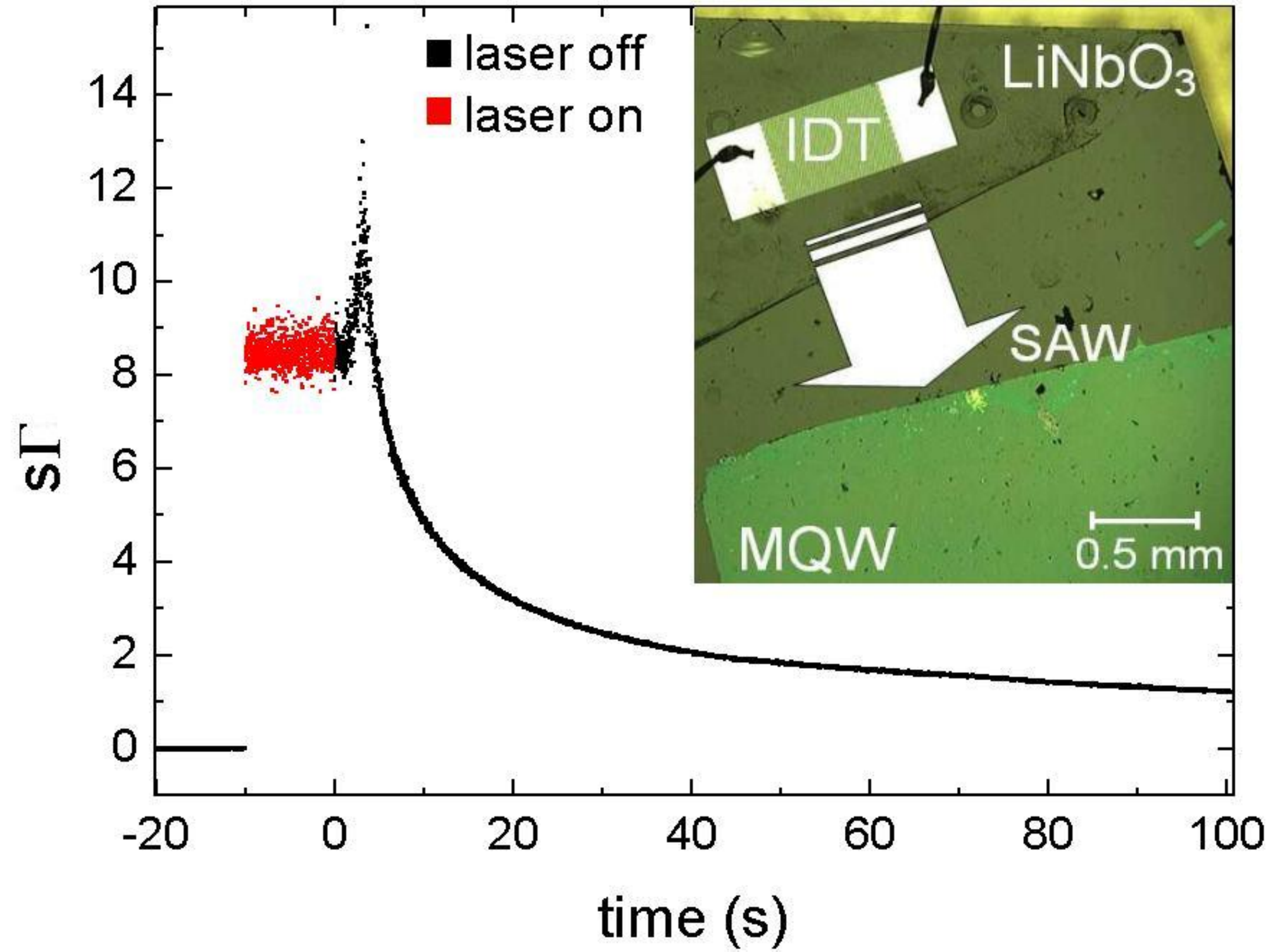}
	\caption{SAW attenuation of the third SAW harmonic ($f_{3}= 345$ MHz) at $T= 35$ K, before, during and after optical excitation. Inset: ZnCdSe/ZnSe heterostructure on a LiNbO$_3$ substrate with IDT. (color online)}
	\label{fig:fig1}
\end{figure}
Our sample consists of a ZnCdSe/ZnSe heterostructure grown by molecular beam epitaxy (MBE). Five optically active, 4 nm wide Zn$_{0.9}$Cd$_{0.1}$Se QWs, separated by 8 nm ZnSe barriers are embedded between two 63 nm ZnSe cladding layers.
After growth, the II-VI semiconductor heterostructure was removed from the substrate and transferred onto the strongly piezoelectric LiNbO$_3$ chip using an epitaxial lift-off method \cite{liftoff, fuhrmann}. The transferred film is positioned between two IDTs on the host LiNbO$_3$ substrate. Due to the increased coupling coefficient, $K_{\rm eff}$, of LiNbO$_3$ compared to Zn(Cd)Se this hybrid sample layout efficiently gives access to higher piezoelectric fields in the semiconductor material which e.g.\ was used to study SAW mediated exciton dissociation \cite{fuhrmann}. A microscope image of one part of the sample is shown in the inset to Fig. \ref{fig:fig1}. In the upper part the IDT with an aperture of 0.58 mm allows for the excitation and/or detection of SAW at $f_0= 115$ MHz and $f_3= 3\cdot f_0$. The SAW launched by the IDT propagates through the semiconductor film (length 2.6 mm) to a second IDT on the opposite side. The total length of the SAW delay line is 5.4 mm and the transmitted power was measured using a conventional network analyzer.
The sample itself was mounted on the coldfinger of an optical helium-flow cryostat equipped with high-frequency connections to the IDTs. For photoluminescence experiments carriers were photogenerated by a laser diode (Laser-1, $\hbar\omega_{ex}=3.18$ eV, $P_{max}= 30$ mW) focused to $\sim 500~\mu$m diameter spot using a $5\times$ microscope objective. The emission from the sample was collected by the same objective, sent to a 0.5 m grating monochromator where it is spectrally analyzed and detected by a $l{\rm N_2}$ cooled charge-coupled device. For excitation energy dependent experiments Laser-1 was replaced by white light spectrally filtered by a monochromator. For optical manipulation of the PPC a second laser diode (Laser-2, $\hbar\omega_{ex}= 1.84$ eV, $P_{max}= 35$ mW) was used.

\subsection{SAW on a ZnCdSe/ZnSe-LiNbO$_3$ hybrid}
Surface acoustic waves are attenuated in the presence of free charge carriers close to the sample surface. In a piezoelectric material the periodic mechanical deformation is accompanied by electric fields which induce currents dissipating energy. Whilst for high SAW intensities ($P_{SAW}>$ 20 dBm) a modulation of the conduction and valence band edges leads to a dissociation of optically generated excitons, thus quenching the photoluminescence \cite{fuhrmann}, in the regime of low powers $P_{SAW}= -20$ dBm, used in all  presented measurements, the photoluminescence signal of the QW is not yet altered or quenched. The SAW propagation hence acts only as a weak perturbation of the system and a measurement of the SAW transmission in this power range provides a non-invasive probe of the conductivity. The transmitted intensity between the two IDTs is given by $I =I_0 e^{-s\Gamma}$, with $s$ being the length of the interaction region and $\Gamma$ the absorption coefficient. For a two-dimensional system \cite{PhysRevB.40.7874} and small SAW amplitudes, the SAW has only a negligible effect on the charge density $\rho(\textbf{r},t)$, therefore, the for a given sheet conductivity, $\sigma$, and absorption coefficient, $\Gamma$, can be written as  
\begin{equation}
	\Gamma=\dfrac{K^{2}_{\rm eff}\cdot k}{2} \cdot\dfrac{\dfrac{\sigma}{\sigma_{m}}}{1+\left(\dfrac{\sigma}{\sigma_{m}}\right)^{2}}.
	\label{eq:gamma}
\end{equation}
In this equation $K_{\rm eff}$ denotes the effective coupling constant, a measure of the piezoelectric strength of the material, $k$ the SAW wave vector and $\sigma_m$ the so-called characteristic conductivity. From a closer examination of equation (\ref{eq:gamma}) one can see that the absorption coefficient as a function of conductivity increases linearly for $\sigma<\sigma_{m}$, reaches a maximum at $\sigma=\sigma_m$ and subsequently decreases proportional to $\sigma^{-1}$ with further increasing conductivity. The characteristic conductivity $\sigma_{m}$ is given by $\sigma_{m}=\epsilon_{0}(1+\epsilon_{hyb})v$ and is thus dependent on the speed of sound $v$ in the material along the respective crystal direction and the relative dielectric constant of the hybrid system. $\epsilon_{hyb}$ can be determined for a given layer sequence \cite{simon96} and depends on the SAW wave vector and, therefore, the different harmonics used in our experiments exhibit different values for the characteristic conductivities
\begin{eqnarray*}
	\sigma_{m}(\mathrm{115MHz})&\approx &1.04\cdot 10^{-6}\,\Omega ^{-1},\\
	\sigma_{m}(\mathrm{345MHz})&\approx &8.75\cdot 10^{-7}\,\Omega ^{-1}.
	\label{eq:eqn1}
\end{eqnarray*}
At these critical conductivities pronounced minima in the transmitted SAW power for both frequencies are expected. The small values for $\sigma_{m}$ indicate the high sensitivity of the SAW to measure small conductivities. In the following the measured parameter will be $s\cdot\Gamma$, the length of the interaction region $s$ multiplied by the absorption coefficient, referred to as the SAW attenuation. While the upper bound of $s$ is determined by the size of the epitaxial lift-off film (2.6 mm) the effective value of $s$ over which changes in the conductivity are detected is mainly limited to the excitation spot size of $\sim 0.5$ mm. Since we keep the spot size constant in our experiments $s\cdot\Gamma$ is a direct measure for the conductivity which are linked by equation \eqref{eq:gamma}. 

Figure \ref{fig:fig1} shows a typical measurement of attenuation at the frequency of the third harmonic ($f_{3}= 345$ MHz) at low temperatures ($T= 35$ K) as a function of time. At the onset of illumination ($t=-10$ s) the attenuation instantly rises. Due to the creation of free charge carriers, the transmission of the SAW is attenuated by $\approx$ 35 dB. At $t=0$ s the laser is turned off, after which the attenuation rises to a maximum, before it very slowly decays on a seconds timescale. This behavior is consistent with equation \eqref{eq:gamma}. In particular, at the maximum of the attenuation the conductivity is equal to the characteristic conductivity $\sigma_m$.\\

\begin{figure}
	\includegraphics[width=0.9\columnwidth]{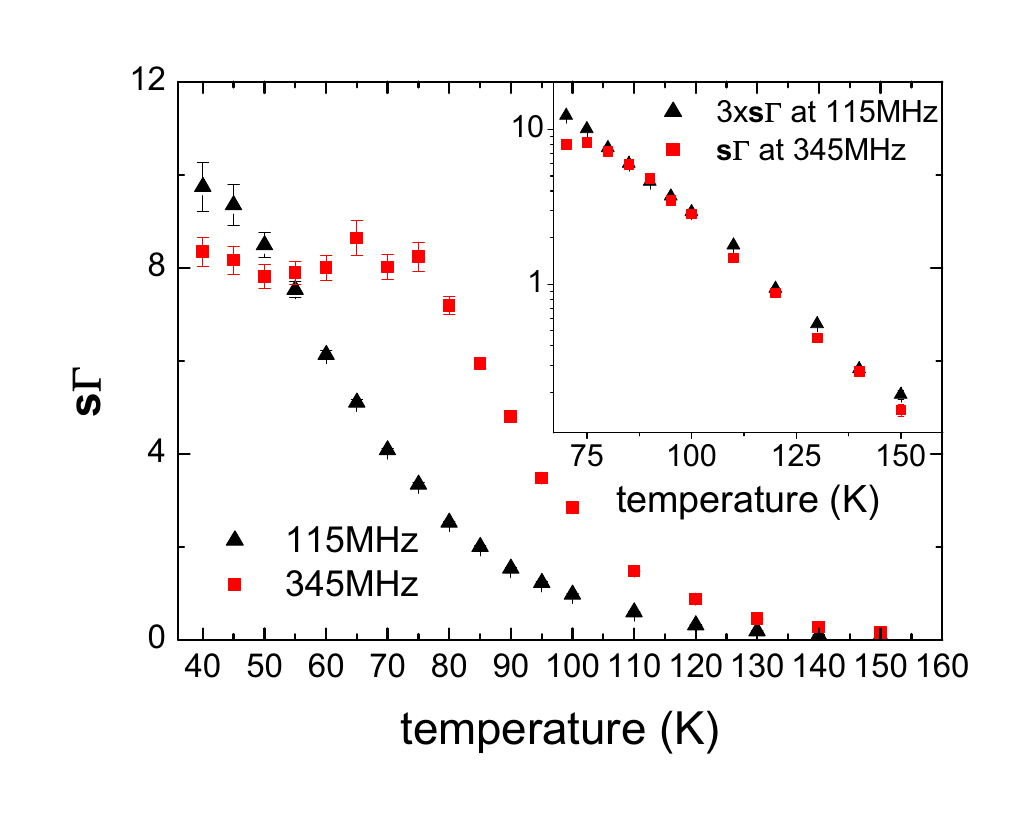}
	\caption{Temperature dependence of the attenuation of the first and third SAW harmonic on illumination. Inset: Measured values of the first harmonic are multiplied by three, to show a $s\Gamma\propto k$ dependence for $\sigma<\sigma_{m}$, i.e. at low temperatures.}
	\label{fig:fig2}
\end{figure}
The increase of the attenuation during illumination as a function of temperature is shown in Fig. \ref{fig:fig2}. For $T>150$ K the transmission of the SAW is not damped and no attenuation is measured. As this is the same temperature above which no photoluminescence could be detected for our sample \cite{fuhrmann}, it hints towards charge carriers from band-to-band transitions as the origin of the PPC. The attenuation of the third harmonic shows a non-linear dependence on the conductivity for $T<70$ K as described by equation (\ref{eq:gamma}). At this temperature the conductivity is very close to the characteristic conductivity $\sigma_{m}\approx 8.75\cdot 10^{-7}\Omega ^{-1}$ of the third harmonic. Since the value of the characteristic conductivity of the first harmonic is higher, the attenuation remains directly proportional to the conductivity over the entire range of temperatures $T>10$ K. Unless stated otherwise we restrict our analysis to data obtained at $f_1$ due to the linear dependency of the conductivity on the SAW attenuation.
For $T>75$ K the measured values of the third SAW harmonic match those of the first multiplied by three (see inset to Fig. \ref{fig:fig2}). This indicates that the slight difference of $\sigma_{m}$(345 MHz) and $\sigma_{m}$(115 MHz) expresses itself in the attenuation for conductivities in the range of $\sigma_{m}$ and shows very nicely that the attenuation is a relative measure of sheet conductivity.

\subsection{Persistent photoconductivity in ZnCdSe quantum wells}

As shown in Fig. \ref{fig:fig1}, the SAW propagation is considerably affected during and long after illumination. This is a rather striking discovery, because in an ideal system in which only charge neutral excitons are generated, the dominant loss mechanism would be radiative recombination. This process takes place on much faster timescales which are not affected by the small amplitudes of the SAW\cite{PhysRevB.57.R6850} used in these experiments. In order to exclude optically active impurities giving rise to levels in the bandgap or surface states, the SAW attenuation was investigated for different photon energies. In this experiment Laser-1 was replaced by white light which was spectrally filtered by our monochromator.
\begin{figure}
	\includegraphics[width=0.9\columnwidth]{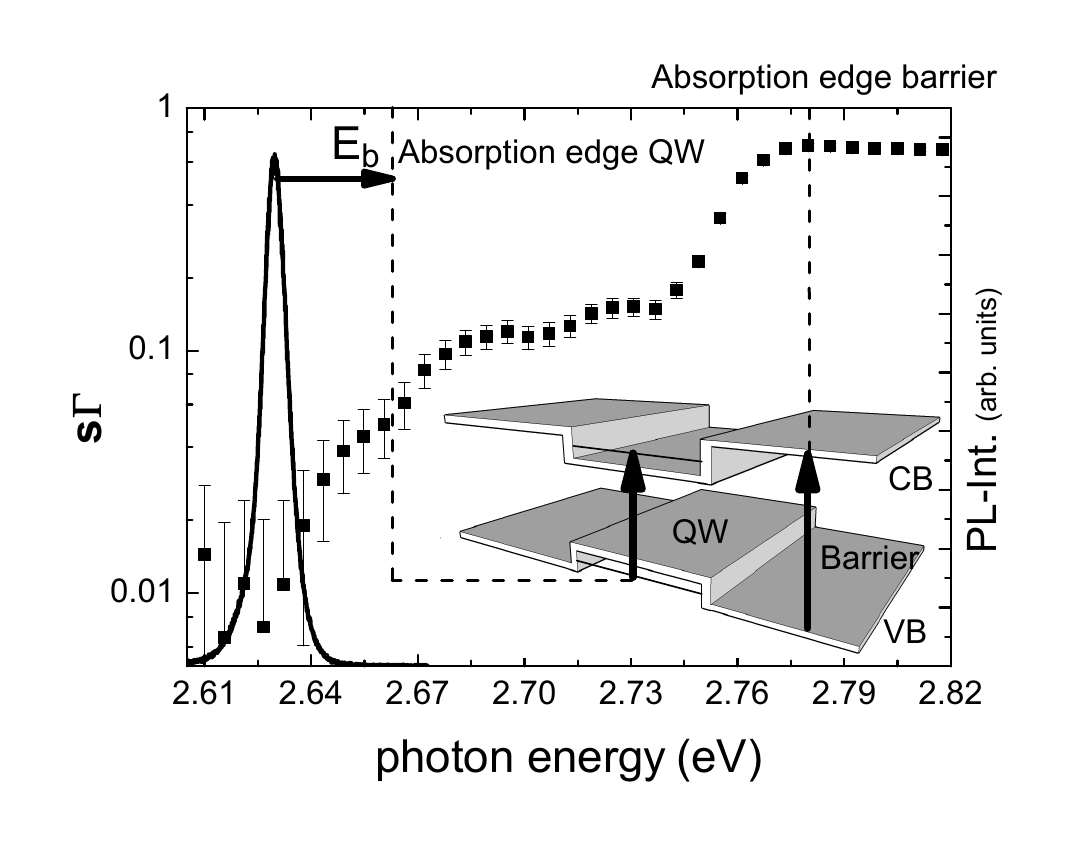}
	\caption{SAW-attenuation and free exciton PL-intensity at $T= 60$ K. The energy of the free exciton recombination and the exciton binding energy $E_{b}\approx$ 32.5 meV \cite{PhysRevB.51.4699} determine the onset of free charge carrier generation in the QW. The energy for the absorption edge of ZnSe is approx. 2.78 eV at 60K \cite{PhysRev.156.850}. The inset shows a schematic of the conduction band (CB) and valence band (VB) of ZnCdSe-QW and ZnSe-barrier in order to illustrate the origin of the attenuating charge carriers (not to scale).}
	\label{fig:fig3}
\end{figure}
Fig. \ref{fig:fig3} shows the SAW attenuation of the first harmonic during illumination at $T= 60$ K as a function the photon energy. At this temperature the attenuation is directly proportional to the conductivity, as indicated by the temperature dependence of the first SAW harmonic shown in Fig. \ref{fig:fig2} and, moreover, the low optical powers used in this experiment ensure that the conductivity is much less $\sigma_m$. Photon energies less or equal than the free exciton recombination energy (a PL spectrum is shown in the figure for comparison), $\hbar\omega\approx$ 2.63 eV, have no effect on the SAW transmission. This excludes activation of defects or impurities with levels within the band gap as the mechanism responsible for the observed effect as it was observed in a previous study on II-VI QWs grown by MBE \cite{Chang}. For energies larger than the QW emission two regions can be distinguished: (i) For energies ranging between 2.63 eV and $\sim2.78$ eV charge carriers are photogenerated solely in the ZnCdSe-QW. (ii) For energies exceeding the bandgap of the ZnSe barrier material, $E_g(\mathrm{ZnSe})\approx 2.78$ eV \cite{PhysRev.156.850}, carriers are also generated outside the QW region. The increase of the signal at this absorption edge indicates that some of the carriers excited in the barrier material relax into the QWs levels.

\begin{figure}
	\includegraphics[width=0.9\columnwidth]{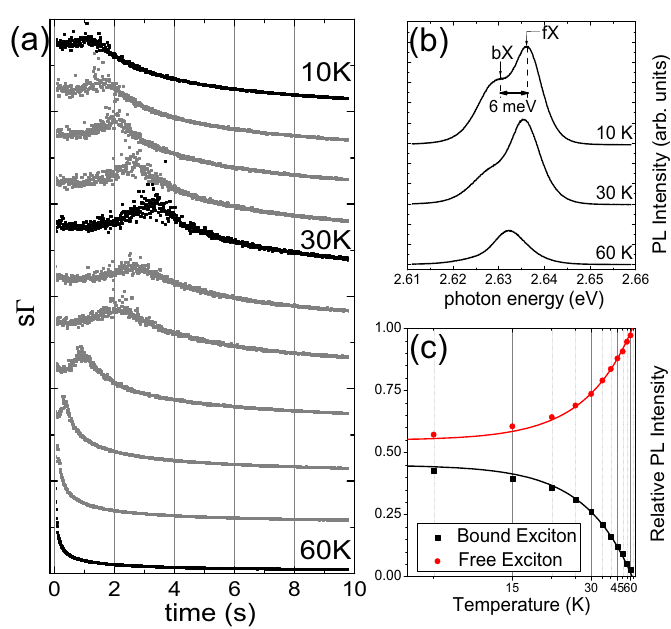}
	\caption{(a) Decay of attenuation of the third SAW harmonic for temperatures $T= 10-60$ K ($\Delta T=5$ K). (b) PL spectra at $T=$ 10, 35 and 60 K showing bound exciton (bX) and free exciton (fX) luminescence. (c) Relative PL intensity of free and bound exciton (symbols) and fit of equations \eqref{eq:fit1} and \eqref{eq:fit2} (solid lines).}
	\label{fig:fig6}
\end{figure}

To obtain further insight in the underlying mechanism we performed temperature dependent SAW-attenuation and PL experiments under illumination by Laser-1 which are summarized in Fig. \ref{fig:fig6}. In the SAW-attenuation experiments the third harmonic was used [Fig. \ref{fig:fig6} (a)] and the laser was switched off at $t = 0$ s. Clearly, the time transients undergo a pronounced change as the temperature is raised. In particular the peak in the SAW-attenuation at which the conductivity becomes comparable to $\sigma_m$ shifts in time and its temporal width changes. At a temperature of $T\sim 30-40$ K both the delay and width of this conductivity peak are maximum. For lower and higher temperatures this peak vanishes and, in addition, a slow and fast decay of the attenuation signal are observed, respectively. Since the observed maximum is linked  to a shift of $\sigma$ from values larger to values smaller than $\sigma_m$ its width and contrast reflect the recombination mechanism and rate of the charge carriers in the system. Though $\sigma (t=0$ s) is not the same for all temperatures the temporal width of the attenuation peak is directly proportional to the recombination rate and these measurements suggest a minimum recombination rate at $T\sim 30$ K. Thus a critical temperature $T_C$ is exceeded in the range between for $25<T<35$ K above which the recombination rate increases rapidly with temperature and exhibits an activated temperature dependence.
This critical temperature corresponds to a thermal energy  of $k_BT_C=2.6$ meV which suggests an activation energy in the same range. Since the excitation energy dependence of the PPC indicates that the underlying mechanism is governed by carriers photogenerated across the bandgap we can use PL spectroscopy to obtain further information over the same temperature range. In Fig. \ref{fig:fig6}(b) we compare the QW emission at $T=10$, 30 and 60 K which are below, close to and above the temperature at which the transition of the SAW attenuation transients is observed. For clarity the corresponding SAW transients at these temperatures are highlighted in Fig. \ref{fig:fig6}(a). A close examination of the PL signal at $T=10$ K shows two contributions of similar intensity which are split by $\sim 6$ meV. The features at lower and higher energy arise from emission of bound (bX) and free excitons (fX), respectively. Whilst for $T= 30$ K bX is suppressed compared to fX but still well resolved at $T=60$ K only fX emission is detected. For a detailed analysis of this effect in the temperature range between $T=10$ and 60 K we plot the relative intensities of bX and fX given by $I_{rel,\mathrm{bX/fX}}=\dfrac{I_{\mathrm{bX/fX}}}{I_{\mathrm{bX}}+I_{\mathrm{fX}}}$ as symbols in Fig \ref{fig:fig6}(c). This data can be described by a thermally activated conversion process from bound (bX) into free excitons (fX):
\begin{eqnarray}
I_{rel,\mathrm{bX}}(T)&=& \dfrac{I_\mathrm{bX}}{I_{\mathrm{bX}}+I_{\mathrm{fX}}}\nonumber\\ ~&=&1-\left[\exp\left({\dfrac{-\Delta E}{k_BT}}\right)+\alpha\right]\label{eq:fit1}\\
I_{rel,\mathrm{fX}}(T)&=& \dfrac{I_\mathrm{fX}}{I_{\mathrm{bX}}+I_{\mathrm{fX}}}\nonumber\\~&=&\exp\left({\dfrac{-\Delta E}{k_BT}}\right)+\alpha .
\label{eq:fit2}
\end{eqnarray}
Here, $\Delta E$ is the exciton localization energy and $\alpha$ is a constant which reflects the number of free excitons for $T\rightarrow 0 $ K for a fixed generation rate i.e.\ optical pump power. The corresponding fits of equations \eqref{eq:fit1} and \eqref{eq:fit2} are plotted as solid lines in Fig. \ref{fig:fig6}(c). From these fits we obtain a localization energy $\Delta E =  4.3$ meV, a value which is comparable to and consistent with the energy splitting between the bX and fX peaks. Moreover, the thermal energy $k_BT_C=2.6$ meV at which the transition is observed in the SAW-attenuation is also in good agreement. Since both SAW-attenuation and PL spectroscopy show a transition in the same temperature range we assume that also for the PPC measured via the SAW-attenuation localization of carriers is the dominant mechanism.\\

\begin{figure}
	\includegraphics[width=0.9\columnwidth]{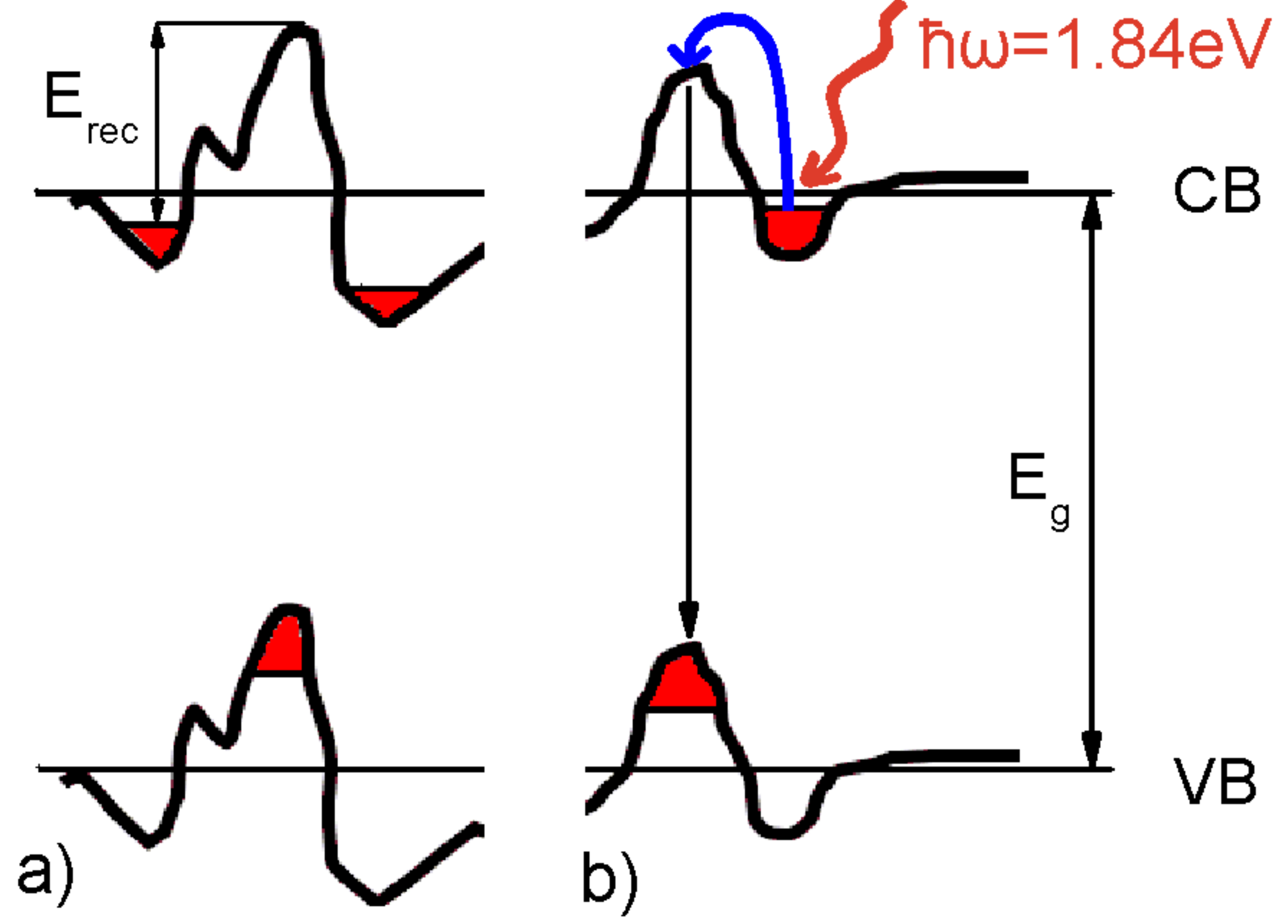}
	\caption{(a) Schematic of spatial modulation of the conduction and valence band edge energies due to RLPFs with an recombination energy $E_{rec}$. Horizontal lines labeled CB and VB show the nominal conduction and valence band edges without RLPF. (b) Excitation of trapped charge carriers using sub-bandgap excitation.}
	\label{fig:fig4}
\end{figure}
In a semiconductor alloy disorder can lead to compositional variations and the resulting variations of the conduction and valence band edges give rise to RLPF. These RLPF can act as local traps for charge carriers as shown schematically in Fig. \ref{fig:fig4}. For our ZnCdSe MQW, local chemical fluctuations i.e.\ of the Cd-content can create potential minima for either electrons or holes, which exhibit an average energy barrier $E_{rec}$ which is a measure for the disorder in the system. Moreover, to transfer carriers between RLPF or release excitons bound to an RLPF the corresponding recombination energy $E_{rec}$ has to be provided for example thermally which gives rise to the observed temperature dependence: For low temperatures charge carriers are localized in the potential minima, thus the resulting conductivity is very low. The separation leads to a reduced wavefunction overlap of electrons and holes inhibiting radiative recombination. This consequently results in an increase of carrier lifetime giving rise to PPC. With increasing thermal energy or temperature, the electron conduction is mediated by hopping transport, since only neighboring minima are accessible. The recombination rate is mainly determined by the distribution of charge carriers in the localization centers, i.e. their wavefunction overlap at the time the illumination is turned off.  At a critical temperature $T_C$, a transition of the conduction mechanism occurs \cite{PhysRevLett.64.2547}: Electrons can now percolate through the network of accessible potential minima, whereas holes remain localized because due to their higher effective mass. The probability for electron trapping is increasing for energetically deeper localization centers where they experience minimal potential energy and the recombination rate reaches a minimum, due to the reduced number of holes in the vicinity. With further increasing temperature, the recombination rate rises until $E_{rec}$ becomes comparable to the thermal energy. From this temperature on PPC breaks down and is no longer observed.\\

As discussed in the previous paragraph, the critical temperature $T_C$ at which the transition of the transport mechanism occurs depends on the depth of the energy minima. These themselves are directly connected to disorder in the material composition and thus the sample quality. Therefore, a small value of $T_C$ indicates that there is a weak and shallow potential modulation due to RLPF. In previous PPC studies on II-VI semiconductor compounds RLPF have been identified in bulk material to be the dominant mechanism at play. The reported values of $T_C>120$ K were found for bulk crystals \cite{Jiang2} and MBE-grown QWs \cite{Chang} . These studies indicate that the introduction of a heterointerface as in a QWs does not necessarily result in a reduction of RLPF. Comparing the findings of reference \cite{Chang} to our work we do not observe PPC effects for sub-bandgap excitation which indicates that RLPF are the dominant mechanism. Moreover, from the presented data we can deduce a value of $T_ C<40$ K for our sample. This strong reduction of the critical temperature and the corresponding shallow band edge modulation of $E_{rec}\lesssim 3$ meV underlines the high quality of our films. This value is comparable to the activation energy between bound and free excitons of $\sim 4.3$ meV. The small deviation could arise from different localization energies for individual carriers and bound electron-hole pairs.

\subsection{Manipulation and saturation of the PPC}
\begin{figure}
	\includegraphics[width=0.9\columnwidth]{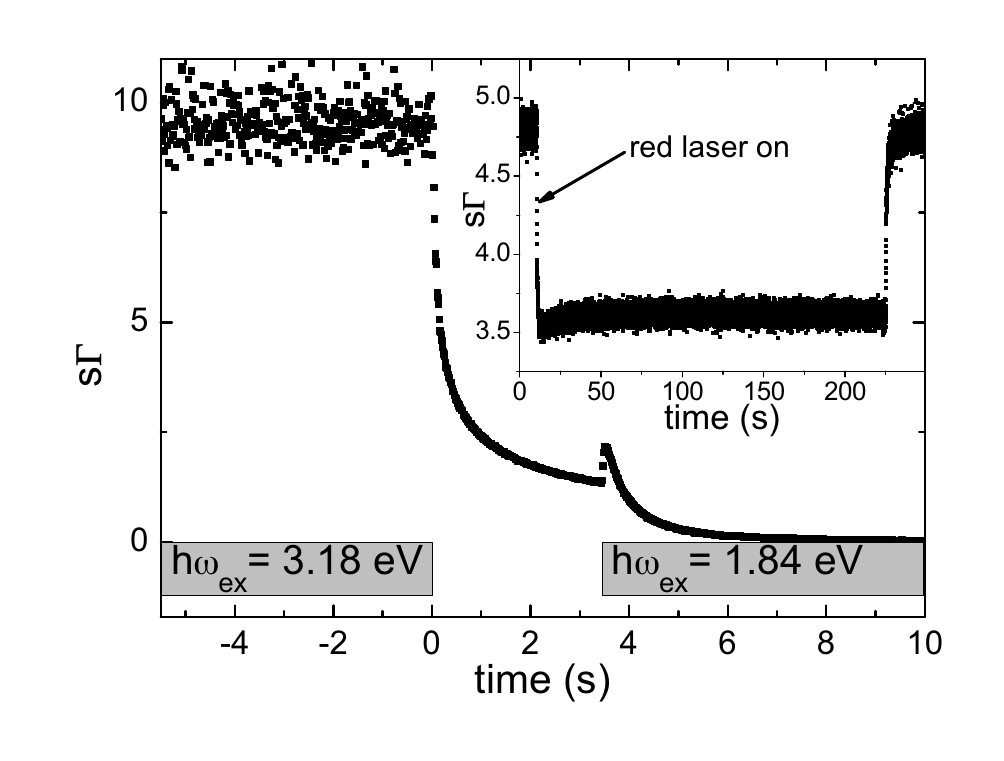}
	\caption{Manipulation of the attenuation after the excitation with above bandgap energy ($T= 45$ K; first SAW harmonic). Irradiation time of the different photon energies is indicated by bars. Inset: Manipulation of the attenuation by optical excitation with $\hbar\omega= 1.84$ eV \emph{during} above bandgap excitation ($T= 60$ K; first SAW harmonic).}
	\label{fig:fig7}
\end{figure}
Another way to transfer carriers trapped in RLPF is by optical excitation using light with an energy larger than $E_{rec}$ as shown schematically in Fig. \ref {fig:fig4} (b). We demonstrated in Fig. \ref{fig:fig3} that for photon energies below the effective bandgap of the QW no extra carriers are generated and, therefore, no PPC can build up. However, if carriers are already present in the system and trapped in RLPF they can be activated giving rise to a response in the PPC. We want to note that the ability to manipulate the PPC is a clear indication that the underlying mechanism in our II-VI QW-system has a different origin e.g.\ in AlGaAs where PPC is dominated by DX-centers. To confirm this assumption a second laser (Laser-2) with an energy $\hbar\omega_{ex}= 1.84$ eV \emph{less} than the effective bandgap of the QW was used for excitation. Due to its smaller energy no excitons are generated by this laser via interband absorption. Fig. \ref{fig:fig7} shows in the main panel the attenuation of the first SAW harmonic over time for a temperature of $T= 45$ K, which ensures a linear dependency on the conductivity. At $t=0$ s the above bandgap excitation $\hbar\omega_{ex}= 3.18$ eV of Laser-1 is turned off (gray shaded area) and the SAW attenuation decays slowly as described before. At $t= 3.5$ s Laser-2 is turned on and a sharp increase of the SAW attenuation is observed. This increase can be understood and explained by an intraband activation of trapped electrons and holes as shown schematically in Fig. \ref{fig:fig4} (b). After activation these carriers now contribute to the sheet conductivity until they recombine. This gives rise to the observed decay of the conductivity even though Laser-2 remains on. This is furthermore consistent with the fact that no additional electrons and holes are photogenerated and only trapped carriers are activated which are lost after recombination. Clearly, the time constant of the decay is faster with Laser-2 switched on which was also observed using an infrared light emitting diode instead of Laser-2. To further confirm this additional activation effect we performed a steady state experiment shown in the inset of Fig. \ref{fig:fig7}. Here, Laser-1 is on all the time and at $t = 10$ s also Laser-2 is used to excite the sample. When Laser-2 is switched on, the SAW attenuation $s\Gamma$ drops by $\sim 25\%$ since additional activation and subsequent recombination of carriers trapped in RPFL occurs. After Laser-2 is switched off again the original attenuation level is recovered.\\

\begin{figure}
	\includegraphics[width=0.9\columnwidth]{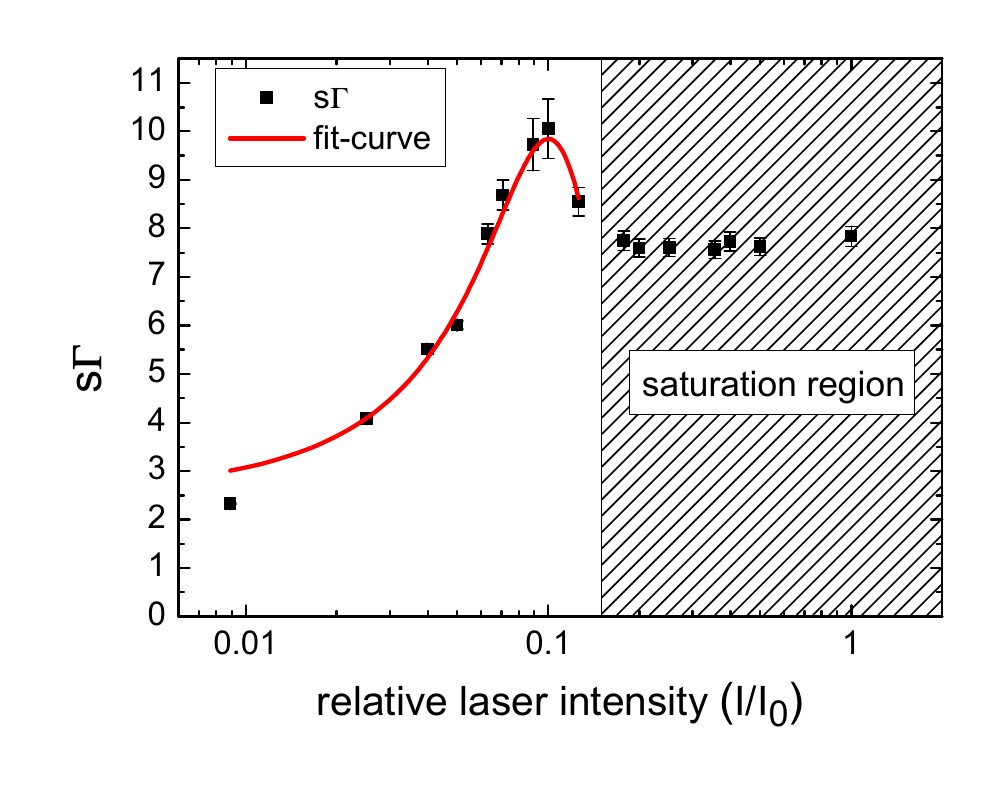}
	\caption{Attenuation of the SAW third harmonic vs. laser intensity at $T= 60$ K. The standard deviation of the attenuation during illumination is indicated by error bars. The solid line represents a fit to the experimental data of equation \eqref{eq:fit-f}.}
	\label{fig:fig5}
\end{figure}
To further support our model based on RLPF we performed SAW-attenuation experiments under continuous excitation using Laser-1 in which the optical pump power was varied and thus, the generation rate of excitons. Assuming that the rates of radiative decay of excitions and recombination of charges trapped in RLPF are independent on the optical pump power, a steady state occupation of RLPF and number of mobile carriers will build up for a given optical power. The measured attenuation as a function of the excitation intensity is shown as symbols in Fig. \ref{fig:fig5}. With increasing laser intensity the SAW attenuation of the third harmonic and, therefore, the sheet conductivity rises and at about ${I}\approx 0.1\cdot I_{0}$, the sheet conductivity matches the characteristic conductivity. The observed behavior can be fitted in this excitation power range using
\begin{equation}
	s\Gamma \left(I/I_{0}\right) = \cfrac{K_{\rm eff}^{2}\cdot s\cdot k}{2}\cfrac{\dfrac{\sigma( I/I_{0})}{\sigma_{m}}}{1+\left(
	\dfrac{\sigma( {I}/{I_{0}})}{\sigma_{m}}\right)^{2}}
	\label{eq:fit-f}
\end{equation}
and assuming for $\sigma$ a single exponential increase given by $\sigma({I}/I_{0})\propto \exp({I}/I_{0})$. Clearly, the result of this fit (solid line in Fig. \ref{fig:fig5}) agrees well over a wide range of intensities. The initial increase of the conductivity can be well understood by assuming a Gaussian variation of the Cd-content from the nominal value of 0.1 and, thus, a Gaussian distribution of $E_{rec}$. Furthermore, this distribution can be correlated with the distribution of the electron kinetic energy in the RLPF: At $T=60$ K the conduction is in the percolation state and, therefore, at low laser intensities the small number of generated electrons is mainly trapped by deep localization centers with an increased Cd-content, minimizing their kinetic energy. With increasing excitation intensity and electron density, electrons have to populate the shallower localization centers with reduced Cd-content. This leads to a strong increase of total kinetic energy of the trapped electrons and explains the exponential rise of sheet conductivity. The overall good agreement between the experimental data and the fit over a wide range of excitation power proves both the validity of equation (\ref{eq:gamma}) and the assumption made for $\sigma({I}/{I_{0}})$. In addition, the observed saturation of the signal is a clear indication to a finite number of localization site available which are fully populated for laser intensities $> 0.2\cdot I_0$.\\

At this point we want to note that the applied SAW-technique is capable to sensitively detect conductivity at very low optical pump powers.  In particular we want to note that the minimum excitation power required to detect PL signals from our samples was in the order of $\sim0.05\cdot I_0$ setting the lower bound at which this invasive method can be applied. In contrast, at this particular pump power already a pronounced SAW-attenuation of $s\Gamma = 6$ is detected. This attenuation level is a factor of 20 - 30 larger than minimum attenuation levels of $0.2\pm0.1$ determined from the baseline and noise in Fig. \ref{fig:fig3}. In addition to this high sensitivity our SAW-attenuation method does not rely on high SAW powers. In particular it is capable to detect in the low SAW power regime where no influence on the PL signal is detected or charge conveyance effects occur.  Thus, this method does not perturb the carrier system in study and resembles a non-invasive probe of the conductivity. In addition, we have shown that despite of the non-linear dependence on the conductivity given by \eqref{eq:gamma} the SAW-response on the conductivity can be analyzed and fitted accurately [Fig. \ref{fig:fig5}] .

\subsection{Conclusions}

In conclusion we studied the attenuation of a SAW in a ZnCdSe/ZnSe-LiNbO$_3$-hybrid. Under optical excitation we observe a variation of the SAW attenuation due to persistent photoconductivity effects in the ternary II-VI alloy. This PPC arises from carriers trapped in and activated from random potential fluctuations due to compositional fluctuations in the material. By performing temperature and excitation energy and power dependent experiments impurity or defect related effects can be excluded. It is shown that the carriers trapped in RLPF can be activated optically and thermally. In the temperature dependent experiments a clear transition from a hopping to a percolation regime is observed which also manifests itself in a suppression of bound exciton emission. We find that the critical temperature below which localization occurs is reduced from $T=120$ K for bulk crystals to $T<40$ K in high qualitiy QWs grown by MBE. The corresponding shallow potential modulation of $E_{rec}=\lesssim 3$ meV underlines the high structural quality achievable in MBE-grown II-VI heterostructures. Furthermore, direct comparison between the applied techniques demonstrates that compared to PL spectroscopy the  SAW-based method yields a very versatile, non-invasive tool for sensitive conductivity measurements which can be applied to other material systems than the one studied in this work.\\

\noindent This work was supported in parts by the German federal government as part of the Cluster of Excellence \emph{Nanosystems Initiative Munich} (NIM).
\bibliography{references.tex}

\begin{thebibliography}{25}
\expandafter\ifx\csname natexlab\endcsname\relax\def\natexlab#1{#1}\fi
\expandafter\ifx\csname bibnamefont\endcsname\relax
  \def\bibnamefont#1{#1}\fi
\expandafter\ifx\csname bibfnamefont\endcsname\relax
  \def\bibfnamefont#1{#1}\fi
\expandafter\ifx\csname citenamefont\endcsname\relax
  \def\citenamefont#1{#1}\fi
\expandafter\ifx\csname url\endcsname\relax
  \def\url#1{\texttt{#1}}\fi
\expandafter\ifx\csname urlprefix\endcsname\relax\def\urlprefix{URL }\fi
\providecommand{\bibinfo}[2]{#2}
\providecommand{\eprint}[2][]{\url{#2}}

\bibitem[{\citenamefont{Wixforth et~al.}(1989)\citenamefont{Wixforth, Scriba,
  Wassermeier, Kotthaus, Weimann, and Schlapp}}]{PhysRevB.40.7874}
\bibinfo{author}{\bibfnamefont{A.}~\bibnamefont{Wixforth}},
  \bibinfo{author}{\bibfnamefont{J.}~\bibnamefont{Scriba}},
  \bibinfo{author}{\bibfnamefont{M.}~\bibnamefont{Wassermeier}},
  \bibinfo{author}{\bibfnamefont{J.~P.} \bibnamefont{Kotthaus}},
  \bibinfo{author}{\bibfnamefont{G.}~\bibnamefont{Weimann}}, \bibnamefont{and}
  \bibinfo{author}{\bibfnamefont{W.}~\bibnamefont{Schlapp}},
  \bibinfo{journal}{Phys. Rev. B} \textbf{\bibinfo{volume}{40}},
  \bibinfo{pages}{7874} (\bibinfo{year}{1989}).

\bibitem[{\citenamefont{Rotter et~al.}(1999)\citenamefont{Rotter, Kalameitsev,
  Govorov, Ruile, and Wixforth}}]{Rotter_PRL}
\bibinfo{author}{\bibfnamefont{M.}~\bibnamefont{Rotter}},
  \bibinfo{author}{\bibfnamefont{A.~V.} \bibnamefont{Kalameitsev}},
  \bibinfo{author}{\bibfnamefont{A.~O.} \bibnamefont{Govorov}},
  \bibinfo{author}{\bibfnamefont{W.}~\bibnamefont{Ruile}}, \bibnamefont{and}
  \bibinfo{author}{\bibfnamefont{A.}~\bibnamefont{Wixforth}},
  \bibinfo{journal}{Phys. Rev. Lett.} \textbf{\bibinfo{volume}{82}},
  \bibinfo{pages}{2171} (\bibinfo{year}{1999}).

\bibitem[{\citenamefont{Rocke et~al.}(1998)\citenamefont{Rocke, Govorov,
  Wixforth, B\"ohm, and Weimann}}]{PhysRevB.57.R6850}
\bibinfo{author}{\bibfnamefont{C.}~\bibnamefont{Rocke}},
  \bibinfo{author}{\bibfnamefont{A.~O.} \bibnamefont{Govorov}},
  \bibinfo{author}{\bibfnamefont{A.}~\bibnamefont{Wixforth}},
  \bibinfo{author}{\bibfnamefont{G.}~\bibnamefont{B\"ohm}}, \bibnamefont{and}
  \bibinfo{author}{\bibfnamefont{G.}~\bibnamefont{Weimann}},
  \bibinfo{journal}{Phys. Rev. B} \textbf{\bibinfo{volume}{57}},
  \bibinfo{pages}{R6850} (\bibinfo{year}{1998}).

\bibitem[{\citenamefont{{Couto Jr}. et~al.}(2007)\citenamefont{{Couto Jr}.,
  Iikawa, Rudolph, Hey, and Santos}}]{couto_PRL}
\bibinfo{author}{\bibfnamefont{O.~D.~D.} \bibnamefont{{Couto Jr}.}},
  \bibinfo{author}{\bibfnamefont{F.}~\bibnamefont{Iikawa}},
  \bibinfo{author}{\bibfnamefont{J.}~\bibnamefont{Rudolph}},
  \bibinfo{author}{\bibfnamefont{R.}~\bibnamefont{Hey}}, \bibnamefont{and}
  \bibinfo{author}{\bibfnamefont{P.~V.} \bibnamefont{Santos}},
  \bibinfo{journal}{Physical Review Letters} \textbf{\bibinfo{volume}{98}},
  \bibinfo{eid}{036603} (pages~\bibinfo{numpages}{4}) (\bibinfo{year}{2007}),
  \urlprefix\url{http://link.aps.org/abstract/PRL/v98/e036603}.

\bibitem[{\citenamefont{B\"{o}defeld et~al.}(2006)\citenamefont{B\"{o}defeld,
  Ebbecke, Toivonen, Sopanen, Lipsanen, and Wixforth}}]{boedefeld_PRB}
\bibinfo{author}{\bibfnamefont{C.}~\bibnamefont{B\"{o}defeld}},
  \bibinfo{author}{\bibfnamefont{J.}~\bibnamefont{Ebbecke}},
  \bibinfo{author}{\bibfnamefont{J.}~\bibnamefont{Toivonen}},
  \bibinfo{author}{\bibfnamefont{M.}~\bibnamefont{Sopanen}},
  \bibinfo{author}{\bibfnamefont{H.}~\bibnamefont{Lipsanen}}, \bibnamefont{and}
  \bibinfo{author}{\bibfnamefont{A.}~\bibnamefont{Wixforth}},
  \bibinfo{journal}{Physical Review B (Condensed Matter and Materials Physics)}
  \textbf{\bibinfo{volume}{74}}, \bibinfo{eid}{035407}
  (pages~\bibinfo{numpages}{5}) (\bibinfo{year}{2006}),
  \urlprefix\url{http://link.aps.org/abstract/PRB/v74/e035407}.

\bibitem[{\citenamefont{Gell et~al.}(2008)\citenamefont{Gell, Ward, Young,
  Stevenson, Atkinson, Anderson, Jones, Ritchie, and Shields}}]{gell_APL}
\bibinfo{author}{\bibfnamefont{J.~R.} \bibnamefont{Gell}},
  \bibinfo{author}{\bibfnamefont{M.~B.} \bibnamefont{Ward}},
  \bibinfo{author}{\bibfnamefont{R.~J.} \bibnamefont{Young}},
  \bibinfo{author}{\bibfnamefont{R.~M.} \bibnamefont{Stevenson}},
  \bibinfo{author}{\bibfnamefont{P.}~\bibnamefont{Atkinson}},
  \bibinfo{author}{\bibfnamefont{D.}~\bibnamefont{Anderson}},
  \bibinfo{author}{\bibfnamefont{G.~A.~C.} \bibnamefont{Jones}},
  \bibinfo{author}{\bibfnamefont{D.~A.} \bibnamefont{Ritchie}},
  \bibnamefont{and} \bibinfo{author}{\bibfnamefont{A.~J.}
  \bibnamefont{Shields}}, \bibinfo{journal}{Applied Physics Letters}
  \textbf{\bibinfo{volume}{93}}, \bibinfo{eid}{081115}
  (pages~\bibinfo{numpages}{3}) (\bibinfo{year}{2008}),
  \urlprefix\url{http://link.aip.org/link/?APL/93/081115/1}.

\bibitem[{\citenamefont{{Couto Jr}. et~al.}(2009)\citenamefont{{Couto Jr}.,
  Lazic, Iikawa, Stotz, Jahn, Hey, and Santos}}]{couto_NatPhot}
\bibinfo{author}{\bibfnamefont{O.~D.~D.} \bibnamefont{{Couto Jr}.}},
  \bibinfo{author}{\bibfnamefont{S.}~\bibnamefont{Lazic}},
  \bibinfo{author}{\bibfnamefont{F.}~\bibnamefont{Iikawa}},
  \bibinfo{author}{\bibfnamefont{J.~A.~H.} \bibnamefont{Stotz}},
  \bibinfo{author}{\bibfnamefont{U.}~\bibnamefont{Jahn}},
  \bibinfo{author}{\bibfnamefont{R.}~\bibnamefont{Hey}}, \bibnamefont{and}
  \bibinfo{author}{\bibfnamefont{P.~V.} \bibnamefont{Santos}},
  \bibinfo{journal}{Nat. Photon.} \textbf{\bibinfo{volume}{3}},
  \bibinfo{pages}{645} (\bibinfo{year}{2009}).

\bibitem[{\citenamefont{Fuhrmann et~al.}(2009)\citenamefont{Fuhrmann, Wixforth,
  Curran, Morrod, Prior, Warburton, and Ebbecke}}]{fuhrmann}
\bibinfo{author}{\bibfnamefont{D.~A.} \bibnamefont{Fuhrmann}},
  \bibinfo{author}{\bibfnamefont{A.}~\bibnamefont{Wixforth}},
  \bibinfo{author}{\bibfnamefont{A.}~\bibnamefont{Curran}},
  \bibinfo{author}{\bibfnamefont{J.~K.} \bibnamefont{Morrod}},
  \bibinfo{author}{\bibfnamefont{K.~A.} \bibnamefont{Prior}},
  \bibinfo{author}{\bibfnamefont{R.~J.} \bibnamefont{Warburton}},
  \bibnamefont{and} \bibinfo{author}{\bibfnamefont{J.}~\bibnamefont{Ebbecke}},
  \bibinfo{journal}{Applied Physics Letters} \textbf{\bibinfo{volume}{94}},
  \bibinfo{eid}{193505} (\bibinfo{year}{2009}).

\bibitem[{\citenamefont{Lang et~al.}(1979)\citenamefont{Lang, Logan, and
  Jaros}}]{Lang}
\bibinfo{author}{\bibfnamefont{D.~V.} \bibnamefont{Lang}},
  \bibinfo{author}{\bibfnamefont{R.~A.} \bibnamefont{Logan}}, \bibnamefont{and}
  \bibinfo{author}{\bibfnamefont{M.}~\bibnamefont{Jaros}},
  \bibinfo{journal}{Phys. Rev. B} \textbf{\bibinfo{volume}{19}},
  \bibinfo{pages}{1015} (\bibinfo{year}{1979}).

\bibitem[{\citenamefont{Firszt et~al.}(2004)\citenamefont{Firszt, Meczynska,
  Legowski, Zakrzewski, Strzalkowski, and Wrobel}}]{Firszt}
\bibinfo{author}{\bibfnamefont{F.}~\bibnamefont{Firszt}},
  \bibinfo{author}{\bibfnamefont{H.}~\bibnamefont{Meczynska}},
  \bibinfo{author}{\bibfnamefont{S.}~\bibnamefont{Legowski}},
  \bibinfo{author}{\bibfnamefont{J.}~\bibnamefont{Zakrzewski}},
  \bibinfo{author}{\bibfnamefont{K.}~\bibnamefont{Strzalkowski}},
  \bibnamefont{and} \bibinfo{author}{\bibfnamefont{M.}~\bibnamefont{Wrobel}},
  \bibinfo{journal}{physica status solidi (c)} \textbf{\bibinfo{volume}{1}},
  \bibinfo{pages}{4} (\bibinfo{year}{2004}).

\bibitem[{\citenamefont{Scholl et~al.}(1994)\citenamefont{Scholl,
  Gersch{\"{u}}tz, Sch{\"{a}}fer, Fischer, Waag, and Landwehr}}]{Scholl1994491}
\bibinfo{author}{\bibfnamefont{S.}~\bibnamefont{Scholl}},
  \bibinfo{author}{\bibfnamefont{J.}~\bibnamefont{Gersch{\"{u}}tz}},
  \bibinfo{author}{\bibfnamefont{H.}~\bibnamefont{Sch{\"{a}}fer}},
  \bibinfo{author}{\bibfnamefont{F.}~\bibnamefont{Fischer}},
  \bibinfo{author}{\bibfnamefont{A.}~\bibnamefont{Waag}}, \bibnamefont{and}
  \bibinfo{author}{\bibfnamefont{G.}~\bibnamefont{Landwehr}},
  \bibinfo{journal}{Solid State Communications} \textbf{\bibinfo{volume}{91}},
  \bibinfo{pages}{491 } (\bibinfo{year}{1994}).

\bibitem[{\citenamefont{Sheinkman et~al.}(1976)\citenamefont{Sheinkman, Shik,
  and Poluprovodn.}}]{Sheinkman76}
\bibinfo{author}{\bibfnamefont{M.}~\bibnamefont{Sheinkman}},
  \bibinfo{author}{\bibfnamefont{A.~Y.} \bibnamefont{Shik}}, \bibnamefont{and}
  \bibinfo{author}{\bibfnamefont{F.~T.} \bibnamefont{Poluprovodn.}},
  \bibinfo{journal}{Sov. Phys. Semicond.} \textbf{\bibinfo{volume}{10}},
  \bibinfo{pages}{128} (\bibinfo{year}{1976}).

\bibitem[{\citenamefont{Queisser and Theodorou}(1986)}]{Queisser}
\bibinfo{author}{\bibfnamefont{H.~J.} \bibnamefont{Queisser}} \bibnamefont{and}
  \bibinfo{author}{\bibfnamefont{D.~E.} \bibnamefont{Theodorou}},
  \bibinfo{journal}{Phys. Rev. B} \textbf{\bibinfo{volume}{33}},
  \bibinfo{pages}{4027} (\bibinfo{year}{1986}).

\bibitem[{\citenamefont{Dissanayake et~al.}(1991)\citenamefont{Dissanayake,
  Huang, Jiang, and Lin}}]{PhysRevB.44.13343}
\bibinfo{author}{\bibfnamefont{A.~S.} \bibnamefont{Dissanayake}},
  \bibinfo{author}{\bibfnamefont{S.~X.} \bibnamefont{Huang}},
  \bibinfo{author}{\bibfnamefont{H.~X.} \bibnamefont{Jiang}}, \bibnamefont{and}
  \bibinfo{author}{\bibfnamefont{J.~Y.} \bibnamefont{Lin}},
  \bibinfo{journal}{Phys. Rev. B} \textbf{\bibinfo{volume}{44}},
  \bibinfo{pages}{13343} (\bibinfo{year}{1991}).

\bibitem[{\citenamefont{Dissanayake et~al.}(1992)\citenamefont{Dissanayake,
  Elahi, Jiang, and Lin}}]{Dissanayake}
\bibinfo{author}{\bibfnamefont{A.}~\bibnamefont{Dissanayake}},
  \bibinfo{author}{\bibfnamefont{M.}~\bibnamefont{Elahi}},
  \bibinfo{author}{\bibfnamefont{H.~X.} \bibnamefont{Jiang}}, \bibnamefont{and}
  \bibinfo{author}{\bibfnamefont{J.~Y.} \bibnamefont{Lin}},
  \bibinfo{journal}{Phys. Rev. B} \textbf{\bibinfo{volume}{45}},
  \bibinfo{pages}{13996} (\bibinfo{year}{1992}).

\bibitem[{\citenamefont{Jiang and Lin}(1989)}]{Jiang2}
\bibinfo{author}{\bibfnamefont{H.~X.} \bibnamefont{Jiang}} \bibnamefont{and}
  \bibinfo{author}{\bibfnamefont{J.~Y.} \bibnamefont{Lin}},
  \bibinfo{journal}{Phys. Rev. B} \textbf{\bibinfo{volume}{40}},
  \bibinfo{pages}{10025} (\bibinfo{year}{1989}).

\bibitem[{\citenamefont{Jiang and Lin}(1990)}]{PhysRevLett.64.2547}
\bibinfo{author}{\bibfnamefont{H.~X.} \bibnamefont{Jiang}} \bibnamefont{and}
  \bibinfo{author}{\bibfnamefont{J.~Y.} \bibnamefont{Lin}},
  \bibinfo{journal}{Phys. Rev. Lett.} \textbf{\bibinfo{volume}{64}},
  \bibinfo{pages}{2547} (\bibinfo{year}{1990}).

\bibitem[{\citenamefont{Jiang et~al.}(1991)\citenamefont{Jiang, Brown, and
  Lin}}]{Jiang1}
\bibinfo{author}{\bibfnamefont{H.~X.} \bibnamefont{Jiang}},
  \bibinfo{author}{\bibfnamefont{G.}~\bibnamefont{Brown}}, \bibnamefont{and}
  \bibinfo{author}{\bibfnamefont{J.~Y.} \bibnamefont{Lin}},
  \bibinfo{journal}{Journal of Applied Physics} \textbf{\bibinfo{volume}{69}},
  \bibinfo{pages}{6701} (\bibinfo{year}{1991}).

\bibitem[{\citenamefont{Jiang et~al.}(1992)\citenamefont{Jiang, Dissanayake,
  and Lin}}]{PhysRevB.45.4520}
\bibinfo{author}{\bibfnamefont{H.~X.} \bibnamefont{Jiang}},
  \bibinfo{author}{\bibfnamefont{A.}~\bibnamefont{Dissanayake}},
  \bibnamefont{and} \bibinfo{author}{\bibfnamefont{J.~Y.} \bibnamefont{Lin}},
  \bibinfo{journal}{Phys. Rev. B} \textbf{\bibinfo{volume}{45}},
  \bibinfo{pages}{4520} (\bibinfo{year}{1992}).

\bibitem[{\citenamefont{Lin and Jiang}(1990)}]{PhysRevB.41.5178}
\bibinfo{author}{\bibfnamefont{J.~Y.} \bibnamefont{Lin}} \bibnamefont{and}
  \bibinfo{author}{\bibfnamefont{H.~X.} \bibnamefont{Jiang}},
  \bibinfo{journal}{Phys. Rev. B} \textbf{\bibinfo{volume}{41}},
  \bibinfo{pages}{5178} (\bibinfo{year}{1990}).

\bibitem[{\citenamefont{Chang et~al.}(2001)\citenamefont{Chang, Yang, Chen,
  Chang, and Chen}}]{Chang}
\bibinfo{author}{\bibfnamefont{C.~W.} \bibnamefont{Chang}},
  \bibinfo{author}{\bibfnamefont{H.~C.} \bibnamefont{Yang}},
  \bibinfo{author}{\bibfnamefont{C.~H.} \bibnamefont{Chen}},
  \bibinfo{author}{\bibfnamefont{H.~J.} \bibnamefont{Chang}}, \bibnamefont{and}
  \bibinfo{author}{\bibfnamefont{Y.~F.} \bibnamefont{Chen}},
  \bibinfo{journal}{Journal of Applied Physics} \textbf{\bibinfo{volume}{89}},
  \bibinfo{pages}{3725} (\bibinfo{year}{2001}).

\bibitem[{\citenamefont{Balocchi et~al.}(2005)\citenamefont{Balocchi, Curran,
  Graham, Bradford, Prior, and Warburton}}]{liftoff}
\bibinfo{author}{\bibfnamefont{A.}~\bibnamefont{Balocchi}},
  \bibinfo{author}{\bibfnamefont{A.}~\bibnamefont{Curran}},
  \bibinfo{author}{\bibfnamefont{T.~C.~M.} \bibnamefont{Graham}},
  \bibinfo{author}{\bibfnamefont{C.}~\bibnamefont{Bradford}},
  \bibinfo{author}{\bibfnamefont{K.~A.} \bibnamefont{Prior}}, \bibnamefont{and}
  \bibinfo{author}{\bibfnamefont{R.~J.} \bibnamefont{Warburton}},
  \bibinfo{journal}{Applied Physics Letters} \textbf{\bibinfo{volume}{86}},
  \bibinfo{eid}{011915} (\bibinfo{year}{2005}).

\bibitem[{\citenamefont{Simon}(1996)}]{simon96}
\bibinfo{author}{\bibfnamefont{S.~H.} \bibnamefont{Simon}},
  \bibinfo{journal}{Phys. Rev. B} \textbf{\bibinfo{volume}{54}},
  \bibinfo{pages}{13878} (\bibinfo{year}{1996}).

\bibitem[{\citenamefont{Liaci et~al.}(1995)\citenamefont{Liaci, Bigenwald,
  Briot, Gil, Briot, and Cloitre}}]{PhysRevB.51.4699}
\bibinfo{author}{\bibfnamefont{F.}~\bibnamefont{Liaci}},
  \bibinfo{author}{\bibfnamefont{P.}~\bibnamefont{Bigenwald}},
  \bibinfo{author}{\bibfnamefont{O.}~\bibnamefont{Briot}},
  \bibinfo{author}{\bibfnamefont{B.}~\bibnamefont{Gil}},
  \bibinfo{author}{\bibfnamefont{N.}~\bibnamefont{Briot}}, \bibnamefont{and}
  \bibinfo{author}{\bibfnamefont{T.~A.} \bibnamefont{Cloitre}},
  \bibinfo{journal}{Phys. Rev. B} \textbf{\bibinfo{volume}{51}},
  \bibinfo{pages}{4699} (\bibinfo{year}{1995}).

\bibitem[{\citenamefont{Hite et~al.}(1967)\citenamefont{Hite, Marple, Aven, and
  Segall}}]{PhysRev.156.850}
\bibinfo{author}{\bibfnamefont{G.~E.} \bibnamefont{Hite}},
  \bibinfo{author}{\bibfnamefont{D.~T.~F.} \bibnamefont{Marple}},
  \bibinfo{author}{\bibfnamefont{M.}~\bibnamefont{Aven}}, \bibnamefont{and}
  \bibinfo{author}{\bibfnamefont{B.}~\bibnamefont{Segall}},
  \bibinfo{journal}{Phys. Rev.} \textbf{\bibinfo{volume}{156}},
  \bibinfo{pages}{850} (\bibinfo{year}{1967}).

\end{thebibliography}
\end{document}